# Topics Emerged in the Biomedical Field and Their Characteristics

Kun Lu[1]

Email: kunlu@ou.edu

School of Library and Information Studies, University of Oklahoma, 401 West Brooks St., Norman, Oklahoma, United States, 73019

Guancan Yang (Email: yanggc@ruc.edu.cn)

Xue Wang (Email: wangxue1120@ruc.edu.cn)

School of Information Resource Management, Renmin University of China, 59 Zhongguancun Street, Haidian District, Beijing, China, 100872

[1] Corresponding author

**Abstract**

This study aims to reveal what kind of topics emerged in the biomedical domain by retrospectively analyzing newly added MeSH (Medical Subject Headings) terms from 2001 to 2010 and how they have been used for indexing since their inclusion in the thesaurus. The goal is to investigate if the future trend of a new topic depends on what kind of topic it is without relying on external indicators such as growth, citation patterns, or word co-occurrences. This topic perspective complements the traditional publication perspective in studying emerging topics. Results show that topic characteristics, including topic category, clinical significance, and if a topic has any narrower terms at the time of inclusion, influence future popularity of a new MeSH. Four emergence trend patterns are identified, including emerged and sustained, emerged not sustained, emerged and fluctuated, and not yet emerged. Predictive models using topic characteristics for emerging topic prediction show promise. This suggests that the characteristics of topics and domain should be considered when predicting future emergence of research topics. This study bridges a gap in emerging topic prediction by offering a topic perspective and advocates for considering topic and domain characteristics as well as economic, medical, and environmental impact when studying emerging topics in the biomedical domain.

**Introduction**

Scientific domains constantly innovate and introduce new topics. However, not all new topics will become popular. And among those that become popular, not all will sustain the popularity. Understanding what topics emerged or not emerged from scientific domains helps us to foresee future fate of a new topic, which has implications for domain researchers, information professionals, and science policy makers. Existing studies on emerging topics generally approach the problem from the publication perspective, that is, to retrieve a set of publications (e.g., research articles or patents) from one or more databases over a period of time, and then try to identify emerging topics from the dataset based on some pre-defined criteria, such as novelty, growth, impact, etc. This study approaches the problem from a topic perspective by tracing a set of new topics in a field to examine the characteristics of the topics that emerged versus not emerged.

Knowledge development involves recognizing new concepts, and studying their properties and relation with existing concepts. Although new concepts are generally first reported using natural language in scientific literature, it is challenging to trace the development in natural language due to terminology variations. Manually curated controlled vocabularies offer a way to study the newly added concepts and how they are discussed in the literature after inclusion. The biomedical field maintains a controlled vocabulary, Medical Subject Headings (MeSH), and uses the vocabulary to index biomedical literature in the PubMed database. The management of the MeSH thesaurus and the indexing process is carried out by professionals with domain knowledge and indexing training. Studying what new terms are introduced to MeSH and how popular they become provides insights into emerging topics in the field. This study retrospectively analyzed newly added MeSH terms from 2001 to 2010, and traced their use for indexing over time. The goal is to understand the characteristics of topics that emerged versus not emerged in the field, and identify emergence trend patterns of topics over time. This study complements existing studies on identifying and predicting emerging topics by revealing the characteristics of emerged topics in a specific domain and identifying trend patterns. In particular, the analysis of the fate of newly added MeSH terms allows the comparison between emerged topics versus the counter-factual samples that have the potential but failed to emerge, which has been noted as lacking in existing studies on this topic (Rotolo et al., 2015). Differing from previous studies on identifying and predicting emerging topics using indicators such as growth, citation patterns, or co-word analysis, this study attempts to examine the influence of topic characteristics on their emergence. This helps explain what drives the growth and impact of emerging topics. The topic characteristics are inherent in the topics themselves and rooted in the domain knowledge. The topic perspective also has the advantage to trace the entire lifecycle of a topic after its introduction, while the publication perspective only reveals what happens in the selected period. The findings from this study will shed light on emerging topics in the biomedical field, and also have implications for more domain-specific approaches to emerging topic prediction.

More specifically, the study attempts to address the following research questions:

*1. What new MeSH concepts were added from 2001 to 2010, and how popular they became? What are the characteristics of those became popular versus those did not?*

*2. What trend patterns emerge? What is the distribution of the MeSH terms over different trend patterns? Do the trend patterns depend on the topic characteristics?*

*3. To which extent can the characteristics of topics predict their future emergence? What factors influence the future emergence of a new MeSH concept?*

## Literature Review
### Emerging topic identification and prediction

Studying trends in research and attempting to predict what is upcoming have attracted attention from the scientometric and research policy communities. In early 2010s, the Intelligence Advanced Research Projects Activity (IARPA) initiated a FUSE (Foresight and Understanding from Scientific Exposition) program aiming to develop automated methods to predict emerging technical capabilities using information in published scientific, technical, and patent literature. The program has boosted the interest in identifying and predicting emerging topics. Many novel methods, especially text-based, have been developed from participants and their collaborators from this program (Babko-Malaya et al., 2015; McKeown et al., 2016). Cozzens et al. (2010) reviewed the concept of emerging technologies and quantitative approaches for monitoring emerging technologies, and recommended combining quantitative and qualitative assessment. A more recent review on emerging research topics divided the research on this topic into three stages: the emergence stage (1965 – 1974), the exploration stage (1974 – 2015), and the development stage (2015 – 2019) (Xu, et al., 2020). The emergence stage is characterized by de Solla Price (1965) and Small (1973) in exploring "research fronts" with citation and co-citation analysis. The exploration stage spans over forty years in exploring and improving methods, mostly bibliometric methods, in characterizing, detecting, and identifying research fronts or emerging topics. The development stage is attributed to be started by Roto, Hicks, and Martin (2015) that reviewed the concept of "emerging technology" and characterized it with five properties: radical novelty, relatively fast growth, coherence, prominent impact, and uncertainty and ambiguity. Although they used "emerging technology" as the terminology, their review covered studies on emerging scientific fields, and emerging research topics. The latest development stage is also characterized by increasing use of large-scale machine learning methods.

*Growth in keyword frequencies*

A number of methods have been proposed to identify emerging topics based on rapid increase in keyword frequencies. This kind of methods generally identify emergent terms.

For example, Chen (2006) delineated the concepts of research front and intellectual base where a research front is the state-of-the-art of a specialty and an intellectual base is what is cited by the research front. He then operationalized research front as a group of words and phrases with sharp growth rate of frequencies. Therefore, the research front terms do not have to be novel. Ohniwa, Hibino and Takeyasu (2010) also used MeSH terms to identify emerging topics. Instead of tracing new MeSH terms, they traced all MeSH terms and identified the ones with top 5% increment in frequency as emerging keywords in a year. Then, the emerging keywords were clustered using co-word analysis to show emerging topics. However, as also pointed out in the original paper, their method not only identified emerging topics, but also included old topics that were revitalized because they did not restrict to only new MeSH terms. Nevertheless, it may be of interest to identify revitalized old topics as well. In addition to sudden increase in keyword frequencies, Guo, Weingart and Börner (2011) also looked at increase in the number of new authors, and the interdisciplinarity of cited references to identify emerging topics. Tu and Seng (2012) combined both novelty and growth in emerging topic detection. Asooja et al. (2016) attempted to predict future TF-IDF of terms using previous scores. Carley et al. (2018) and Porter et al. (2019) proposed an emergence indicator that considers the novelty, growth, persistence, and community of terms for detecting emergent terms. The emergent terms were then used to identify emergent authors, organizations, and countries.

*Document clusters*

A different approach to emerging topic identification relies on various clustering methods, including citation, co-citation, bibliographic coupling, and hybrid. This approach generally identifies clusters of documents as emerging topics. Compared with emergent terms, document clusters need further interpretation and tend to be more general.

Upham and Small (2010) considered clusters of highly cited papers as research fronts and treated new research fronts as emerging research areas. Glänzel and Thijs (2012) combined bibliographic coupling and textual similarity to generate document clusters for each time slice. Cross-citations between clusters in different time slices were used to identify emerging topics. Small, Boyack and Klavans (2014) combined direct citation and co-citation clusterings to nominate clusters that are new and rapid growing. The emerging topics are defined as clusters of documents that are new and grow rapidly, which can be rather broad. However, it suits their goal to develop a global method that can analyze an entire citation database, such as Scopus or Web of Science. Wang (2018) adapted the definition of emerging technologies in Rotolo et al. (2015) to define emerging research topics, and operationalized the criteria of an emerging research topic: radical novelty, fast growth, coherence, and scientific impact. A direct-citation-based clustering method was used to identify research topics, and emerging topics were selected based on these four criteria. A more recent study used global models of science that are derived from direct-citation-based clusters of all publications indexed by the Scopus database in the selected time periods (Klavans, Boyack & Murdick, 2020). The focus of the study is on predicting future growth, so the topics do not have to be novel.

*Machine learning*

As noted in Xu et al. (2020), machine learning has been increasingly used in emerging topic identification since 2015. Both unsupervised and supervised machine learning have been applied. Some studies used unsupervised machine learning, topic models, to generate candidate topics, and then assessed their emergence. Others have applied supervised machine learning to identify emerging research topics from learned discriminative models on designed features.

For example, Ranaei and Suominen (2017) used Latent Dirichlet Allocation and Dynamic Topic Modeling to identify emerging technologies from vehicle related patent data. Xu, et al. (2021) developed a framework for detecting and forecasting emerging technologies based on topic models. Kyebambe, et al. (2017) relied on supervised machine learning to train a model from automatically labeled emerged technologies to identify new emerging technologies on patent data. Features used to discriminate emerging technologies from non-emerging ones include: number of claims, number of citations, number of citations to non-patent literature, technology cycle time, patent class, cited technology similarity index, and cited patents assignee similarity index. Lee et al. (2018) used 18 patent indicators to train neural network models to predict emerging technologies in early stages of technology development.

As the task of emerging topic identification becomes more clearly defined and more methods are proposed, comparative studies on different approaches have emerged. A recent study by Ranaei et al. (2020) evaluated three approaches for identifying emerging technologies, including TF-IDF, the emergence score (Carley et al., 2018), and Latent

Dirichlet Allocation (LDA). They found all three methods are able to track emergent aspects of selected technologies, where TF-IDF can highlight emergent terms at a detailed level, the emergence score allows stability, and LDA is able to provide context for easy interpretation.

**Emergence trend patterns**

A majority of studies have focused on detecting or identifying emerging topics. Some studies have also showed the trend patterns of emerging topics. For example, Wang (2018) showed annual publication trends for the emerging topics identified using her method. However, she did not further characterize the emergence patterns. Lee et al. (2018) characterized the emergence trends using two indicators: emergingness and trend, where emergingness represents the present development of a technology and trend measures the future trend of a technology. This is helpful for technology planning. It is interesting that they also noticed the fluctuations of emergingness. This is different from most existing studies on emerging topic identification, which dichotomize emergingness as either emerging or not emerging. Nevertheless, they did not further examine the emergence patterns. Emergence patterns provide rich information on the process of emergence. In some cases, the trend patterns can show if the emergence has ended or the topic is still emerging.

**Study on new MeSH terms**

Many studies have used MeSH terms for scientometric analysis or information retrieval. However, few studies have specifically focused on newly added MeSH terms, and even fewer have used new MeSH terms to study emerging research topics.

Moerchen et al. (2008) aimed to predict new MeSH terms from the trend of word stems in titles and abstracts of biomedical literature. After identifying emerging MeSH terms using increment ratio, Ohniwa, Hibino and Takeyasu (2010) compared the initial year of a MeSH's appearance with the year that the MeSH is identified as emerging. They found many MeSH terms are considered emerging the year they first appeared. However, the number of such terms decreased from mid-90s. Tsatsaronis et al. (2013) developed a temporal classifier model to predict which MeSH terms will be expanded in the future. This is not a direct study on new MeSH terms, but a study on where the new MeSH terms will be added. McCray and Lee (2013) analyzed the evolution of terms in the "F" section of MeSH from 1963 to 2008. They found changes in the MeSH section were driven by both internal considerations (e.g., developing a more principled ontological structure) and external forces (e.g., development of biomedical knowledge). Balogh et al. (2019) studied the preference of attachment and detachment of new MeSH terms when they were added to the hierarchical structure. Kastrin and Hristovski (2019) examined the evolution of MeSH co-occurrence networks over time by looking at new MeSH terms each year and the edge formation processes. They found most new edges are formed between old MeSH terms. It should be noted that the new MeSH terms in their study are not the newly added ones to the MeSH thesaurus, but the new MeSH terms entered co-occurrence network before a specific year.

A brief review of existing literature suggests most studies on emerging topic identification approach the problem from the publication perspective, and focus on developing indicators or features to identify/predict emergence. Few studies have examined emergence trend patterns, and very few have approached the problem from the topic perspective using new MeSH terms. This study aims to bridge this gap in the literature.

## Method

### Data collection

This study uses a similar data collection method from previous work (Lu, 2020), but automates the data collection process and extends the coverage to include new MeSH introduced from 2001 to 2010. First, the lists of new MeSH terms were retrieved from ftp://nlmpubs.nlm.nih.gov/online/mesh/1999-2010/newterms/. Each term on the lists was looked up in the MeSH RDF linked data representation (https://id.nlm.nih.gov/mesh/). If a term has since been deleted from MeSH, it was excluded. If a term has a "Previously Indexing" term, it was excluded because this indicates the MeSH term is an existing concept indexed under a different name, while this study aims to track new concepts. Then, a search was performed using NCBI E-utilities with the appropriate PubMed query language to "Restrict to MeSH Major Topic" and "Do not include MeSH terms found below this term in the MeSH hierarchy" to search the publications in PubMed that are indexed with the subject term but not its narrower terms. The first article indexed with the MeSH term was identified. If the publication date is more than five years before the subject term was added, the subject term was excluded because this indicates an existing concept. The MeSH categories of V (Publication Characteristics) and Z (Geographicals) are not used to characterize subject content according to MeSH record types (https://www.nlm.nih.gov/mesh/intro_record_types.html), and thus were excluded from further analysis.

For the remaining MeSH terms, searches were then performed in PubMed for articles that have been indexed with the MeSH terms by the end of 2019. The numbers of articles indexed by the MeSH terms are used as a measure of their popularity. Articles indexed by the narrower concepts of the selected MeSH terms were not included because the study aims to reveal the popularity of the selected terms not their narrower concepts. This provides a more precise picture on which concept is popular. Table 1 listed the number of new MeSH from 2001 to 2010 and selected MeSH for this study. A total of 1279 new MeSH terms were selected.

Table 1: The number of new MeSH added and selected each year.

| Year | 2001 | 2002 | 2003 | 2004 | 2005 | 2006 | 2007 | 2008 | 2009 | 2010 |
|---|---|---|---|---|---|---|---|---|---|---|
| # of New MeSH | 184 | 847 | 1250 | 666 | 487 | 943 | 494 | 456 | 446 | 422 |
| # of selected | 20 | 74 | 175 | 69 | 24 | 478 | 144 | 114 | 89 | 92 |

**Topic characteristics**

To understand what kind of topics emerged and what did not, three topic characteristics were examined, including topic category, clinical significance, and whether a new MeSH has any narrower terms at the time of inclusion. These topic characteristics are inherent features of topics.

*Topic category*

Different topic categories may have an influence on the future trend of a new MeSH term. There are 16 main branches in the top-level hierarchy, including Anatomy, Organisms, Diseases, Chemicals and Drugs, Analytical, Diagnostic and Therapeutic Techniques and Equipment, etc (https://www.nlm.nih.gov/bsd/disted/meshtutorial/meshtreestructures/index.html). Popularity of different topic categories was analyzed quantitatively. Further analysis was carried out on popular and unpopular terms in major categories to reveal the characteristics of emerged topics versus not emerged ones in the biomedical field.

*Clinical significance*

Biomedical field has its goal in improving healthcare. Concepts that have clinical significance or usage may attract more attention in the field. This study examines the effect of clinical significance on a topic's future popularity. To ascertain if a MeSH concept has clinical significance/usage, we searched the PubMed using the MeSH term. If the MeSH term is assigned to at least one article with the type of "Clinical Trial", we consider the MeSH concept having clinical significance. It should be noted that although we relied on publications for the clinical significance indicator, this characteristic only depends on the semantics of a MeSH concept and domain knowledge.

*Narrower terms*

The addition of new MeSH terms does not always indicate the discovery of new concepts. Some terms are added due to structural warrant (Svenonius, 1989). For example, the generic term "Primate T-lymphotropic virus 2" was added in 2002 to collocate two narrower concepts "Human T-lymphotropic virus 2" and "Simian T-lymphotropic virus 2". This kind of terms may become less popular because the MeSH indexing policy requires the most specific terms to be used[2]. To examine this effect, the selected MeSH were categorized into those without narrower terms and those with narrower terms at the time of the inclusion. The effect of having narrower terms is then examined on the popularity of the new MeSH concepts.

**Emergence patterns**

Most studies on emerging research topics have focused on methods to identify the topics, and simply categorized topics into emerging or non-emerging topics. However, there can be different patterns of emergence. Some topics sustain popularity after emergence, while others may not. Emergence pattern analysis reveals the popularity over time. The selected MeSH terms were categorized into different trend patterns. Analysis was carried out to reveal the distribution of MeSH terms over different trend patterns.

**Prediction**

Predictors are designed to predict if a MeSH term would become an emerging topic based on the topic characteristics. The selected MeSH terms in each year were divided into four quartiles according to the total number of indexed articles, with Q1 being the least popular and Q4 being the most popular. This ensures the popularity of MeSH terms is only compared with the ones that are added in the same year. The top 25% (Q4) were considered as emerging topics and the other three quartiles as non-emerging topics. Down sampling was applied to the non-emerging topics in the training set for equal training samples (Japkowicz, 2000). Our experiments showed that the down sampling method effectively improves prediction performance.

In logistic regression, we estimate the probability of the $i$th MeSH term would become an emerging topic ($\pi_i$) as a function of predictors. The logistic regression model can be written as follows, where $Z_i$ is a linear combination of the predictors:

$$\pi_i = \frac{e^{z_i}}{1+e^{z_i}} \qquad (1)$$

$$Z_i = \beta_0 + \beta_1 X_{i1} + \beta_2 X_{i2} + \beta_3 X_{i3} + \epsilon_i \qquad (2)$$

where $X_{i1}$ corresponds to the clinical significance of the $i$th topic, $X_{i2}$ measures if the $i$th topic has any narrower terms at the time of inclusion, and $X_{i3}$ measures the topic category of the $i$th topic. Dummy coding was used to encode the topic category variable. It should be noted that we solely relied on topic characteristics to predict emerging topics and did not use any other features because we focus on the effect of topic characteristics alone. Due to the nature of our predictors differing from major predictors for emerging topic prediction, we anticipate that these topic characteristics are complementary to existing predictors.

## Results

**Descriptive analysis**

Table 2 provides the descriptive statistics for the total number of articles indexed by the MeSH terms and the average number of articles indexed per year, which indicates the popularity of the concepts. The distribution of popularity is very skewed and does not conform to normal distribution.

---

[2] https://www.nlm.nih.gov/bsd/disted/meshtutorial/principlesofmedlinesubjectindexing/principles/04.html

Table 2. Descriptive statistics of MeSH popularity (*N*=1279).

|  | Mean | Min. | 1st Quartile | Median | 3rd Quartile | Max. |
|---|---|---|---|---|---|---|
| # of articles indexed | 375.88 | 0 | 36 | 113 | 317.5 | 10926 |
| # of articles indexed per year | 27.28 | 0 | 2.68 | 8.64 | 24.35 | 602.18 |

Table 3 and Table 4 list the most and least popular new MeSH terms and total number of articles indexed until the end of 2019. The least popular ones had not indexed any articles since inclusion. In fact, there are 26 such MeSH terms.

Table 3. The most popular new MeSH terms since inclusion until 2019.

| MeSH term | Total # of articles indexed |
|---|---|
| Recognition (Psychology) | 10926 |
| RNA Interference | 10425 |
| Quantum Dots | 7914 |
| beta Catenin | 7382 |
| Nanocomposites | 7327 |
| STAT3 Transcription Factor | 7059 |
| Hypoxia-Inducible Factor 1, alpha Subunit | 7011 |
| Adiponectin | 6518 |
| Nanotubes | 6314 |
| Metabolome | 5804 |

Table 4. The least popular new MeSH terms since inclusion until 2019.

| MeSH term | Total # of articles indexed |
|---|---|
| 2-Oxoisovalerate Dehydrogenase (Acylating) | 0 |
| Benzaldehyde Dehydrogenase (NADP+) | 0 |
| Endo-1,3(4)-beta-Glucanase | 0 |
| Glycine Dehydrogenase (Decarboxylating) | 0 |
| Kappapapillomavirus | 0 |
| Malate Dehydrogenase (NADP+) | 0 |
| Malonate-Semialdehyde Dehydrogenase (Acetylating) | 0 |
| Methylmalonate-Semialdehyde Dehydrogenase (Acylating) | 0 |
| Nitrate Reductase (NAD(P)H) | 0 |
| Nitrate Reductase (NADH) | 0 |

**Topic characteristics analysis**

*Topic category*

The distribution of topic categories over the quartiles is presented in Table 5. It should be noted that one MeSH term may be assigned multiple topic categories (e.g., "Nanocapsules" belongs to both D - "Chemicals and Drugs" and E - "Techniques and Equipment"). Table 5 considers the popularity of topic categories. The total number of topic category occurrences of the selected MeSH terms is 1306 rather than 1279.

Table 5: Broad category distribution over popularity quartiles (Q1 the least popular, Q4 the most popular, Quartiles are assessed in each year, and then aggregated, $N$=1306).

|  | A | B | C | D | E | F | G | H | I | J | K | L | N |
|---|---|---|---|---|---|---|---|---|---|---|---|---|---|
| **Q1** | 1 | 96 | 6 | 183 | 2 | 1 | 13 | 2 | 1 | 0 | 0 | 0 | 1 |
| **Q2** | 4 | 81 | 22 | 195 | 7 | 0 | 16 | 0 | 2 | 0 | 0 | 1 | 8 |
| **Q3** | 3 | 77 | 10 | 211 | 10 | 1 | 13 | 1 | 1 | 2 | 1 | 0 | 5 |
| **Q4** | 4 | 45 | 7 | 184 | 21 | 4 | 30 | 4 | 3 | 11 | 3 | 3 | 10 |
| **Total** | 12 | 299 | 45 | 773 | 40 | 6 | 72 | 7 | 7 | 13 | 4 | 4 | 24 |

Overall, Chemicals and Drugs (D) accounts for 59.19% of new MeSH terms added from 2001 to 2010, followed by Organisms (B) at 22.89%. These two categories account for 82.08% of all new MeSH terms. Many other categories only have small shares of the new MeSH concepts.

To examine if the popularity of a new MeSH depends on its topic category, a chi-square independence test was conducted. Because many categories only have few new

MeSH terms which will violate the assumption of chi-square test, only categories B, C, D, E, and G are retained for the test. In total, they account for 94.10% of the new concepts. The chi-square test result is statistically significant (d.f.=12, p<0.01). Figure 1 shows the deviation from independence, where boxes rising above the baseline indicate observed frequencies are greater than expected under the assumption of independence, and vice versa. It is observed that Organisms (B) category is more likely to be in the least popular quartile Q1 and less likely in the most popular quartile Q4. Diseases (C) category is more likely to be observed in Q2, a relatively unpopular quartile. Chemicals and Drugs (D) seem to be roughly evenly distributed among quartiles. E (Techniques and Equipment) and G (Phenomena and Processes) are more likely to be in the most popular quartile. This indicates that the popularity of new MeSH terms depends on which broad categories they belong to.

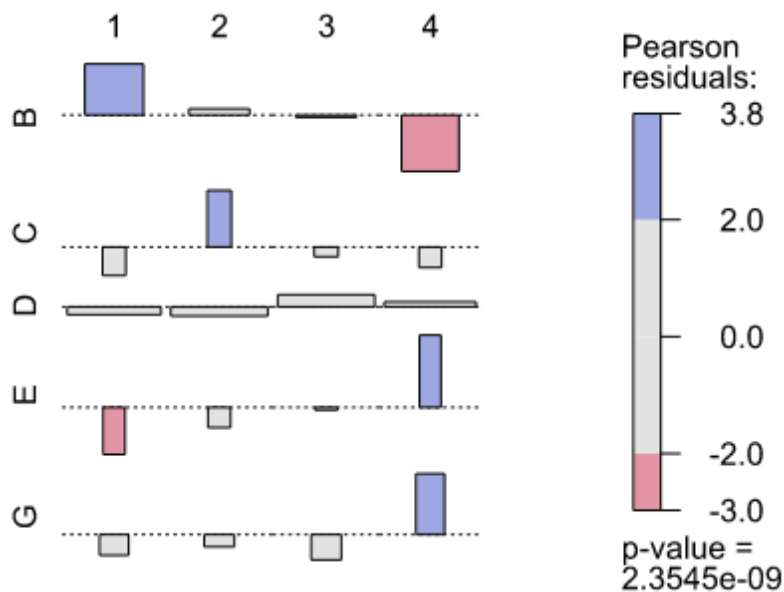

Figure 1: Deviation from independence for topic category distribution over different quartiles.

*Clinical significance*

This section examines the effect of clinical significance on the popularity of a new MeSH concept. Table 6 shows that 614 of the selected MeSH terms were assigned to at least one "Clinical Trial" article, and the rest 665 were not assigned to any "Clinical Trial" articles. Their median popularity differs greatly (272 vs. 45). This difference is statistically significant with a non-parametric Kruskal-Wallis test (d.f.=1, p<0.01). This indicates that MeSH concepts with clinical significance are more popular than those without clinical significance. Figure 2 shows the boxplot of the popularity of the two groups.

Table 6: Popularity of new MeSH with vs. without "Clinical Significance" (*N*=1279).

|  | Median popularity | Number of observations |
|---|---|---|
| **With Clinical Significance** | 272 | 614 |
| **Without Clinical Significance** | 45 | 665 |

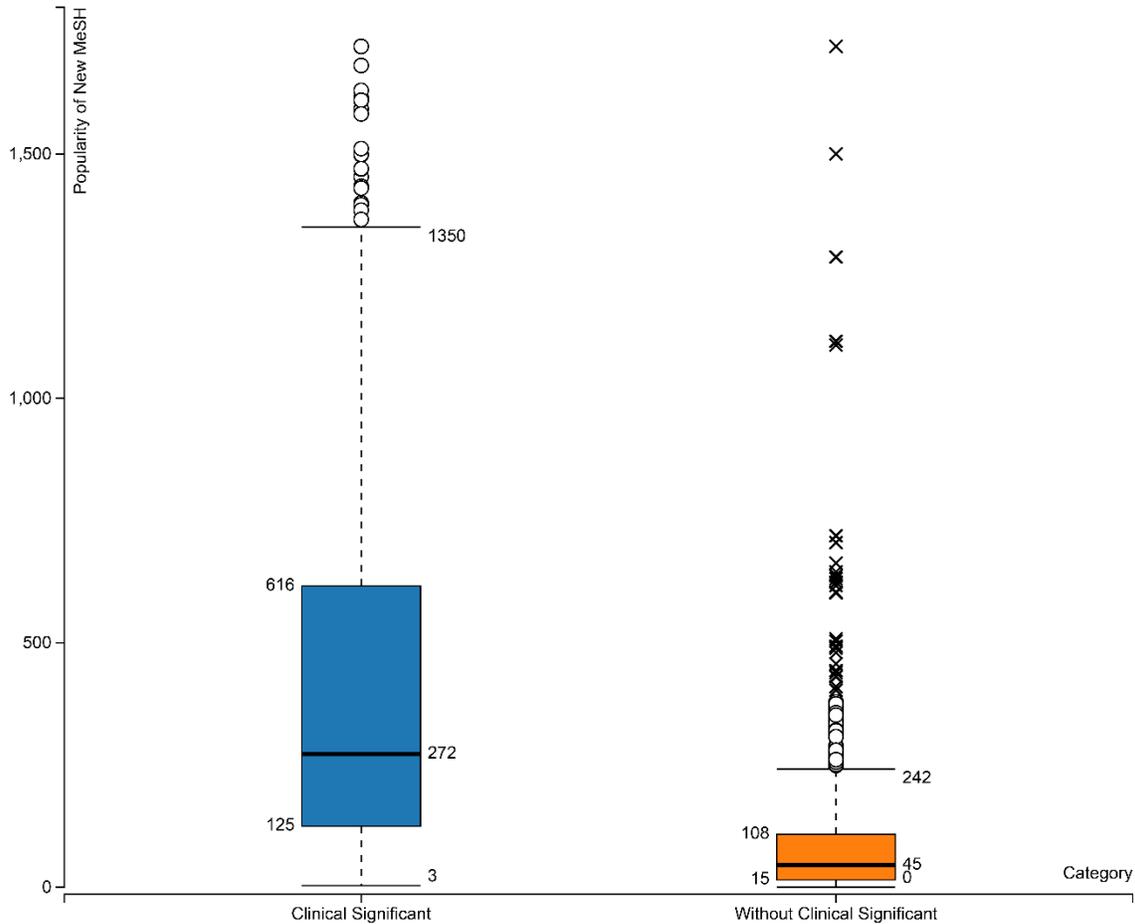

Figure 2: BoxPlot of popularity of selected MeSH terms with or without "Clinical Trial" articles.

*Narrower terms*

    Table 7 shows that the median popularity of new MeSH without narrower terms at the time of addition is 122, much greater than those with narrower terms at 77. The difference is statistically significant with a non-parametric Kruskal-Wallis test (d.f.=1, p<0.01). This indicates the new MeSH without narrower terms at the time of their addition is more popular than those with narrower terms. Figure 3 provides a boxplot of the

popularity of the two groups. It can be observed although the medians are significantly different, the distributions of popularity values of the two groups have large overlap.

Table 7: Popularity of new MeSH with or without narrower terms at the time of addition ($N=1279$).

|  | Median popularity | Number of observations |
|---|---|---|
| **With narrower terms at the time of addition** | 77 | 273 |
| **Without narrower terms at the time of addition** | 122 | 1006 |

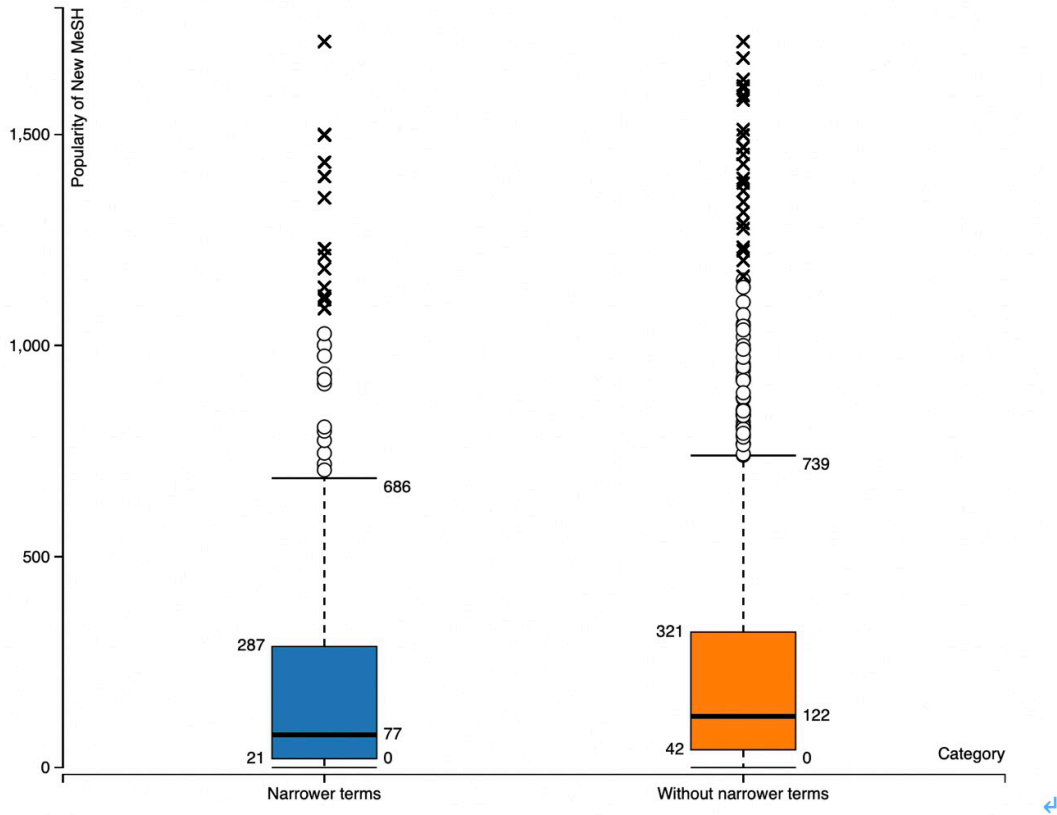

Figure 3: BoxPlot of popularity of selected MeSH terms with or without narrower terms at the time of addition.

*Popular organisms*

Although a large portion of new Organisms added to MeSH did not become popular topics, 45 of them made into the most popular quartile. To examine what features these popular organisms have, Table 8 lists the most popular new Organisms and conjectured reasons for their popularity. A microbiologist and a plant biologist were consulted for the reasons. It is hypothesized that based on the observations in Table 8, new Organisms added to MeSH would become popular if they have environmental, economic (industrial or agricultural), and medical impact.

Table 8: Popular topics belong to Organisms category (One popular concept "Plant Stomata" was excluded because when added to MeSH in 2008, it was assigned to Organisms category. However, it was changed to Antonomy category in 2009 where it should be. Therefore, it was not considered Organisms category in this analysis).

| MeSH term | Category | Total # of articles indexed | Conjectured reasons for popularity |
| --- | --- | --- | --- |
| **Norwalk-like Viruses** | Viruses | 3436 | Pathogen |
| **Endangered Species** | Animals | 2109 | Environmental |
| **Metapneumovirus** | Viruses | 1046 | Pathogen |
| **Rhodobacteraceae** | Bacteria | 719 | Environmental (photosynthesis) |
| **Rhizophoraceae** | Plants | 703 | Environmental |
| **Geobacter** | Bacteria | 663 | Environmental |
| **Salt-Tolerant Plants** | Plants | 604 | Economic (Agricultural) |
| **Vibrio cholerae O1** | Bacteria | 592 | Pathogen |
| **Lactobacillales** | Bacteria | 553 | Food |
| **Nigella sativa** | Plants | 525 | Medical |
| **Chloroflexi** | Bacteria | 480 | Environmental |

Environmental and economic impact of organisms cannot be easily assessed without adequate domain knowledge. Pathogens, on the other hand, are easier to identify in MeSH since their annotation usually indicates the link from organisms to the diseases they cause[3]. For example, in the annotation field of the MeSH term "Norwalk-like Viruses", it states "infection: coord IM with CALICIVIRIDAE INFECTIONS + probably GASTROENTERITIS (IM)". This is a clear indication for a pathogen. In addition, the scope note field sometimes specifies which organisms the infection affects. For example, the scope note field of "Norwalk-like Viruses" states "A genus in the family CALICIVIRIDAE, associated with epidemic gastroenteritis in humans…" With this, we can categorize the Organisms concepts into four categories: Pathogens for human, Pathogens for other organisms, Pathogens for both human and other organisms, and Non-pathogen. Table 9 lists the medians of the popularity of MeSH in the four categories since the data is highly skewed:

---

[3] https://www.nlm.nih.gov/tsd/cataloging/trainingcourses/mesh/mod4_060.html

Table 9: Popularity of different organisms (*N*=299).

|  | **Median** | **Number of Observations** |
|---|---|---|
| **Pathogens for human** | 167 | 7 |
| **Pathogens for other organisms** | 21 | 19 |
| **Pathogens for both human and other organisms** | 31 | 17 |
| **Non-pathogens** | 44 | 256 |

Pathogens for human attract the most attention with a median of 167 publications, which is greater than those of the other three groups. Interestingly, pathogens for other organisms and pathogens for both are less popular than non-pathogens, although not statistically significant. Provided pathogen is one of the factors that influence the popularity of a new organism added to MeSH, non-pathogens may still become popular if they have, for example, environmental or economic impact (see examples in Table 8). Non-parametric Kruskal-Wallis test was used to test the difference among the popularity of four groups of organisms. The difference of popularity among different groups is statistically significant (d.f.=3, $p<0.05$). Figure 4 shows the boxplot of the popularity of different pathogens and non-pathogens.

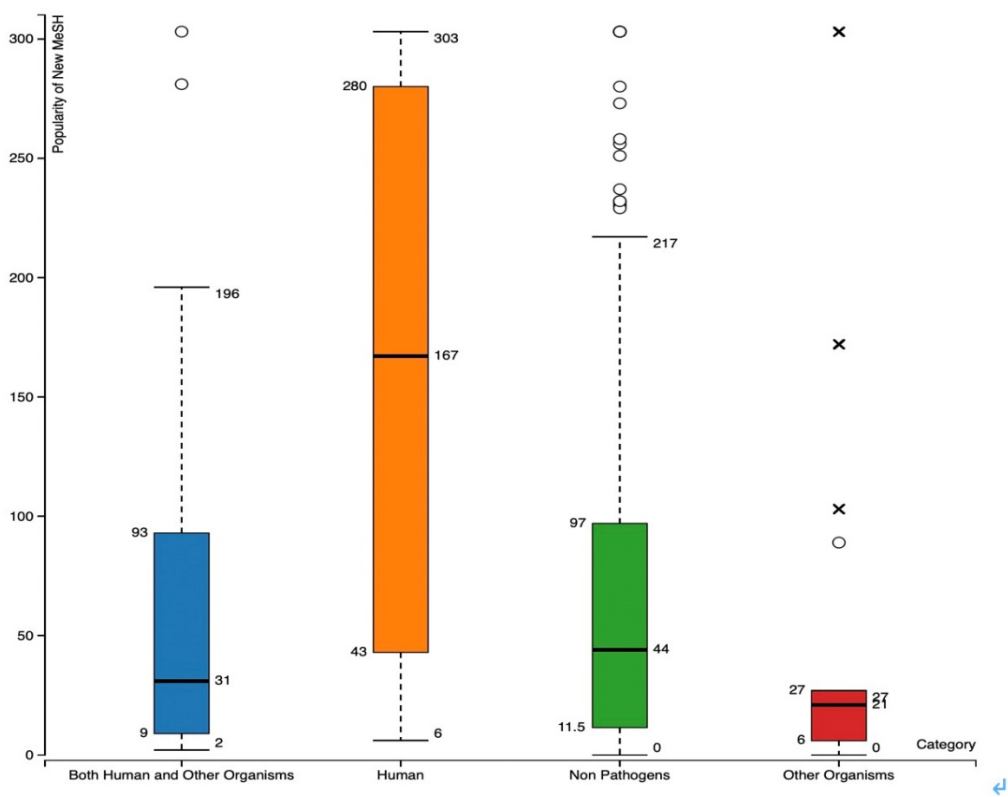

Figure 4: BoxPlot of popularity of different organisms based on pathogen categories.

**Emergence trend patterns**

Not all new MeSH terms show the same emergence trend pattern. Depending on the change in popularity over time, the new MeSH terms can be categorized into four emergence trend patterns: emerged and sustained, emerged but not sustained, emerged and fluctuated, and not yet emerged. The operational definition in this study is: if in any year, the number of articles indexed by a MeSH term reached 25 articles or above (> 3rd Quartile of the average number of articles indexed per year), the MeSH term is considered "Emerged". If in the following years, the number of articles indexed by it dropped below 25 articles for a consecutive of two years (because one year can be by chance), it is considered "not sustained". And if after that, the topic emerged again (reaching 25 articles a year), it is categorized as "Emerged and Fluctuated". Using this criteria, the new MeSH terms are categorized depending on their emergence patterns (Table 10).

Table 10: Trend Category Distribution (N=1279).

| Trend Category | Number of MeSH in the category |
| --- | --- |
| Emerged and Sustained | 315 (24.62%) |
| Emerged not Sustained | 101 (7.90%) |
| Emerged and Fluctuated | 68 (5.32%) |
| Not yet Emerged | 795 (62.15%) |
| **Total MeSH** | **1279** |

It is not surprising that a majority of the new concepts are in the category of "Not yet Emerged" since the threshold for emerging is above the third quartile in a year. Among the ones that emerged, a majority did sustain. However, there are still 101 new MeSH that emerged but not sustained, and 68 emerged and fluctuated, accounting for 20.91% and 14.08% of the emerged topics. The following figures show an example for each of the four categories in Table 10.

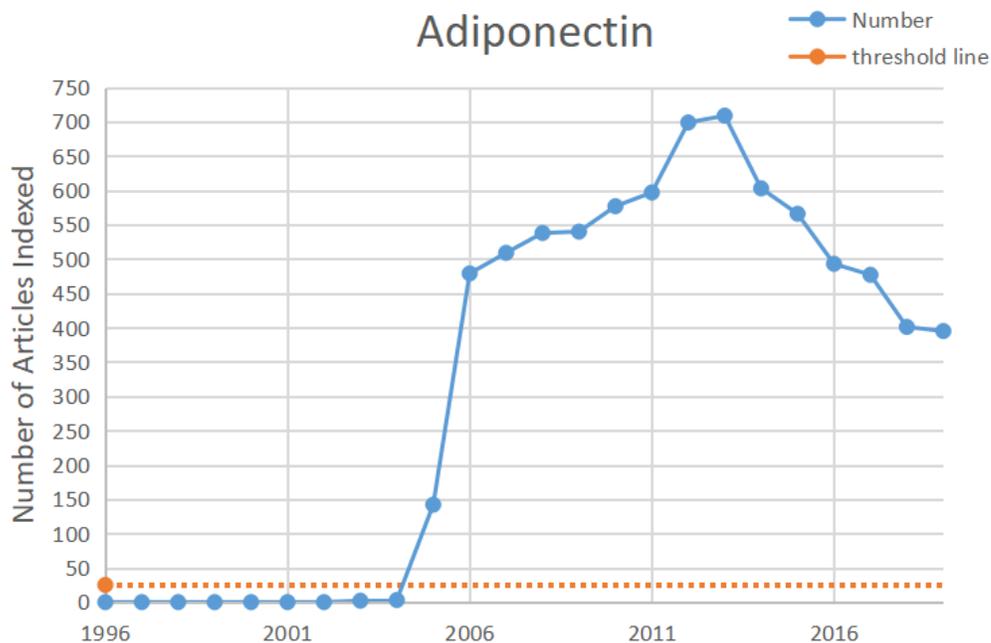

Figure 5: An example for "Emerged and Sustained" trend category.

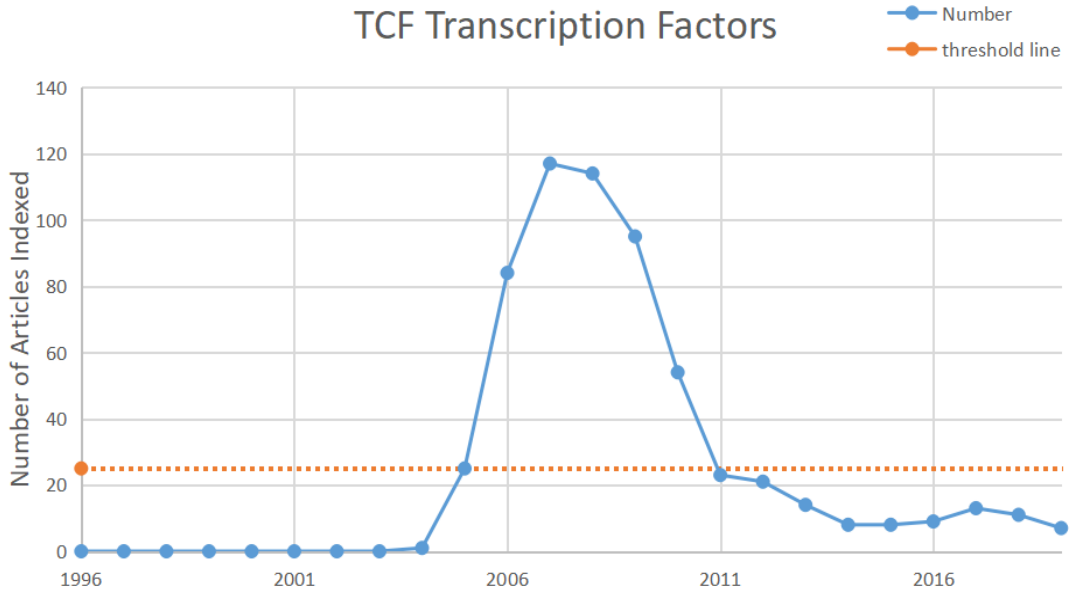

Figure 6: An example for "Emerged not Sustained" trend category.

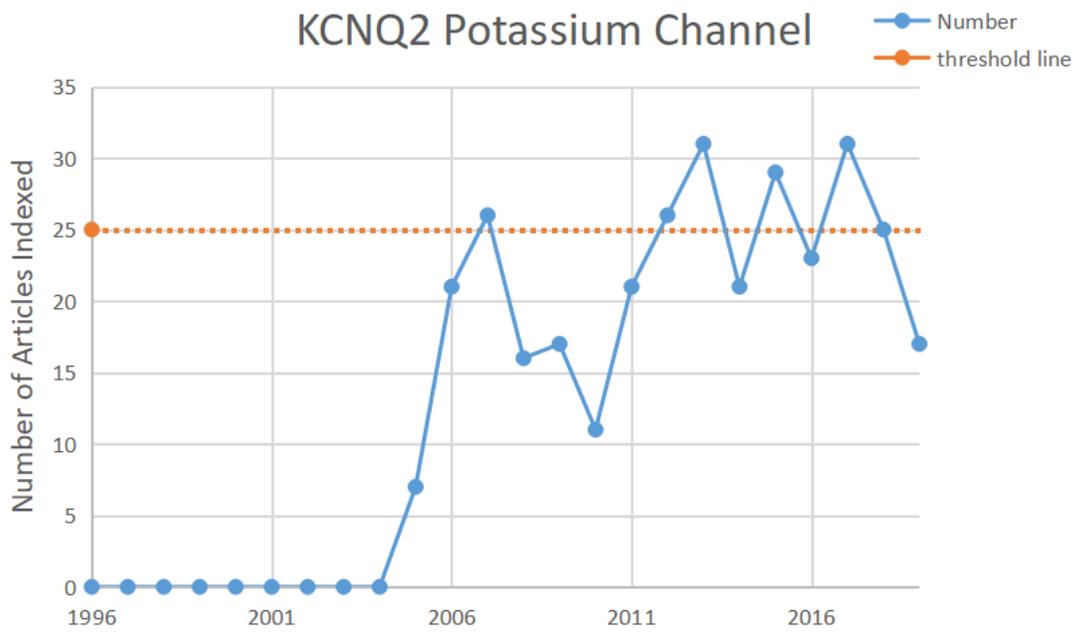

Figure 7: An example for "Emerged and Fluctuated" trend category.

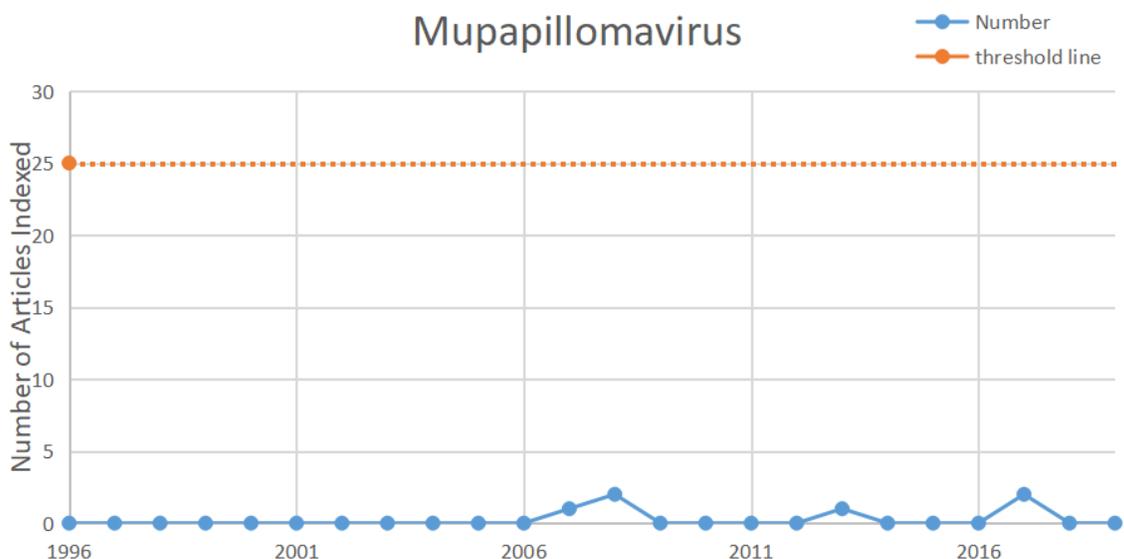

Figure 8: An example for "Not yet Emerged" trend category.

To examine if the topic categories of a MeSH influence its emergence trend pattern, we tallied the frequency distribution of topic categories over trend patterns (Table 11).

Table 11: The frequency distribution of topic categories over trend patterns ($N$=1306).

|  | A | B | C | D | E | F | G | H | I | J | K | L | N | Total |
|---|---|---|---|---|---|---|---|---|---|---|---|---|---|---|
| **Emerged and Sustained** | 4 | 20 | 12 | 211 | 19 | 4 | 27 | 5 | 3 | 11 | 3 | 2 | 8 | 329 |
| **Emerged not Sustained** | 0 | 12 | 2 | 74 | 4 | 1 | 7 | 0 | 1 | 0 | 0 | 0 | 3 | 104 |
| **Emerged and Fluctuated** | 0 | 5 | 0 | 54 | 4 | 0 | 4 | 0 | 0 | 0 | 0 | 1 | 0 | 68 |
| **Not yet Emerged** | 8 | 262 | 31 | 434 | 13 | 1 | 34 | 2 | 3 | 2 | 1 | 1 | 13 | 805 |
| **Total** | 12 | 299 | 45 | 773 | 40 | 6 | 72 | 7 | 7 | 13 | 4 | 4 | 24 | 1306 |

According to Table 11, eleven out of 13 new MeSH in the category of J (Technology, Industry, Agriculture) are considered "Emerged and Sustained". Technology, industry, and agriculture are applications of biomedical research. This suggests new concepts that generate applications are likely to emerge and sustain. Three of the four new MeSH in the topic category K (Humanities) emerged and sustained. They are "September 11 Terrorist Attacks", "Iraq War, 2003 -", and "Afghan Campaign 2001-". It shows that the attention of biomedical research has also been attracted to the three events in terms of the health-related issues there. Two of the four new terms in L (Information Science) made into the category of "Emerged and Sustained", indicating the sustained prevalence of information science methods in assisting biomedical research. Two thirds of the new terms in F (Psychiatry and Psychology) are considered "Emerged and Sustained". This shows the sustained attention to psychological health. Terms in H (Disciplines and Occupations) reflect the goal of MeSH in describing disciplines, but not really about biomedical concepts. Similar can be said to the category I (Anthropology, Education, Sociology, and Social Phenomena) that describes broad connections of biomedical research.

To focus on the majority of new concepts, we selected the topic categories of B, C, D, E, and G. Figure 9 shows that topic categories are associated with trend patterns with a chi-square test of independence (d.f.=12, p<0.01). Topic category B (Organisms) is not likely to emerge or sustain. Topic categories D, E, and G are more likely to emerge and sustain than observed by chance. The "Emerged and Fluctuated" pattern seems to be more likely for D and E, than for G. This suggests the interest on Phenomena and Processes may be more stable than that on Chemicals/Drugs and Techniques/Equipment.

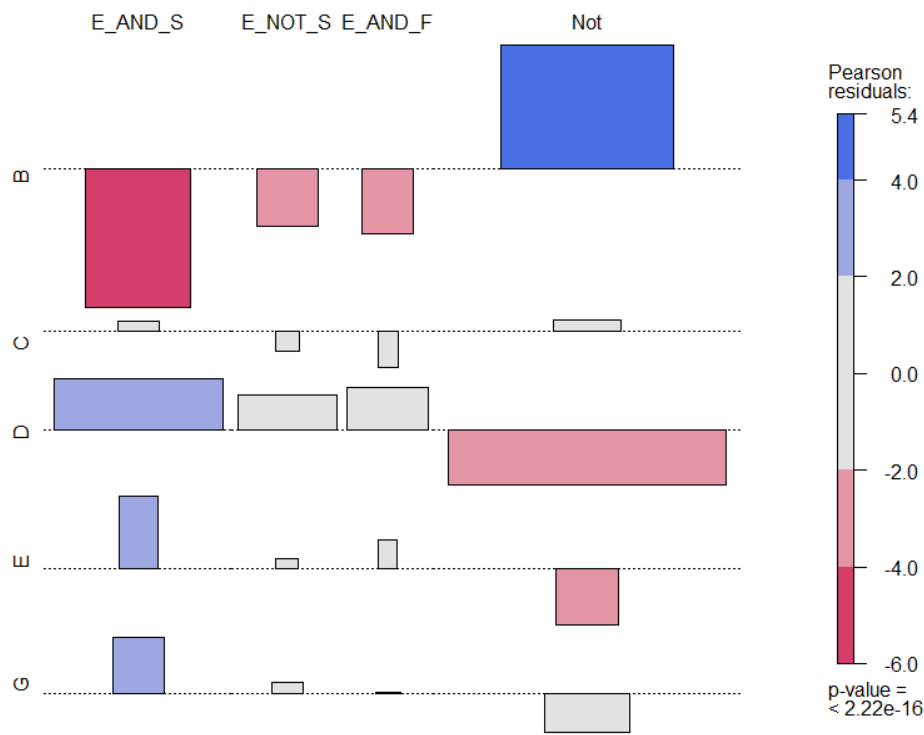

Figure 9: Deviation from independence for topic category distribution over trend patterns.

**Emerging topic prediction based on topic characteristics**

Among the three predictors, topic category and whether a new MeSH has any narrower terms are available at the time of its inclusion, while clinical significance may be known later depending on if a "Clinical Trial" article has been published on this topic. We examined the impact of forecasting time points on the performance of the predictive model by predicting at the $M$th year since a Mesh term is added using only information available at the time, where $M$ ranges from 1 to 10. The performance is evaluated using accuracy, precision, recall, F-measure and the Critical Success Index (CSI) score. CSI is calculated as TP/(TP+FP+FN), where TP is the number of true positive, FP is false positive, and FN is false negative. Logistic regression models were trained to use topic characteristics to predict emerging topics as in equations (1) and (2). Five-fold cross-validation was used to obtain evaluation metrics in Table 12.

Table 12: Performance of predictive models at the $M$th year ($N$=1306).

| Predicted at the $M$th Year | Accuracy | Recall | Precision | F-measure | CSI |
| --- | --- | --- | --- | --- | --- |
| 1 | 0.7418 | 0.3811 | 0.4960 | 0.4310 | 0.2747 |
| 2 | 0.7324 | 0.4512 | 0.4774 | 0.4639 | 0.3020 |
| 3 | 0.7332 | 0.5244 | 0.4818 | 0.5022 | 0.3353 |
| 4 | 0.7426 | 0.6220 | 0.4988 | 0.5536 | 0.3827 |
| 5 | 0.7379 | 0.6860 | 0.4923 | 0.5732 | 0.4018 |
| 6 | 0.7293 | 0.7134 | 0.4815 | 0.5749 | 0.4034 |
| 7 | 0.7167 | 0.7043 | 0.4657 | 0.5607 | 0.3895 |
| 8 | 0.7081 | 0.7591 | 0.4586 | 0.5718 | 0.4003 |
| 9 | 0.6894 | 0.7744 | 0.4402 | 0.5613 | 0.3902 |
| 10 | 0.6933 | 0.7774 | 0.4443 | 0.5654 | 0.3941 |

If the prediction takes place in the first year when a new MeSH is added, the model has an overall accuracy of 74.18%, an F-measure of 43.10% and a CSI of 27.47%. When predicting in the tenth year, the model has an overall accuracy of 69.33%, an F-measure of 56.54% and a CSI of 39.41%. Overall, the recall of the models maintains an upward trend over time, indicating that more emerging topics can be predicted over time. Figure 10 plotted the trend of CSI over the years of forecasting. The model reaches 33.53% at the CSI index when predicting in the third year. The best CSI is achieved when predicting in the sixth years since a new MeSH is introduced.

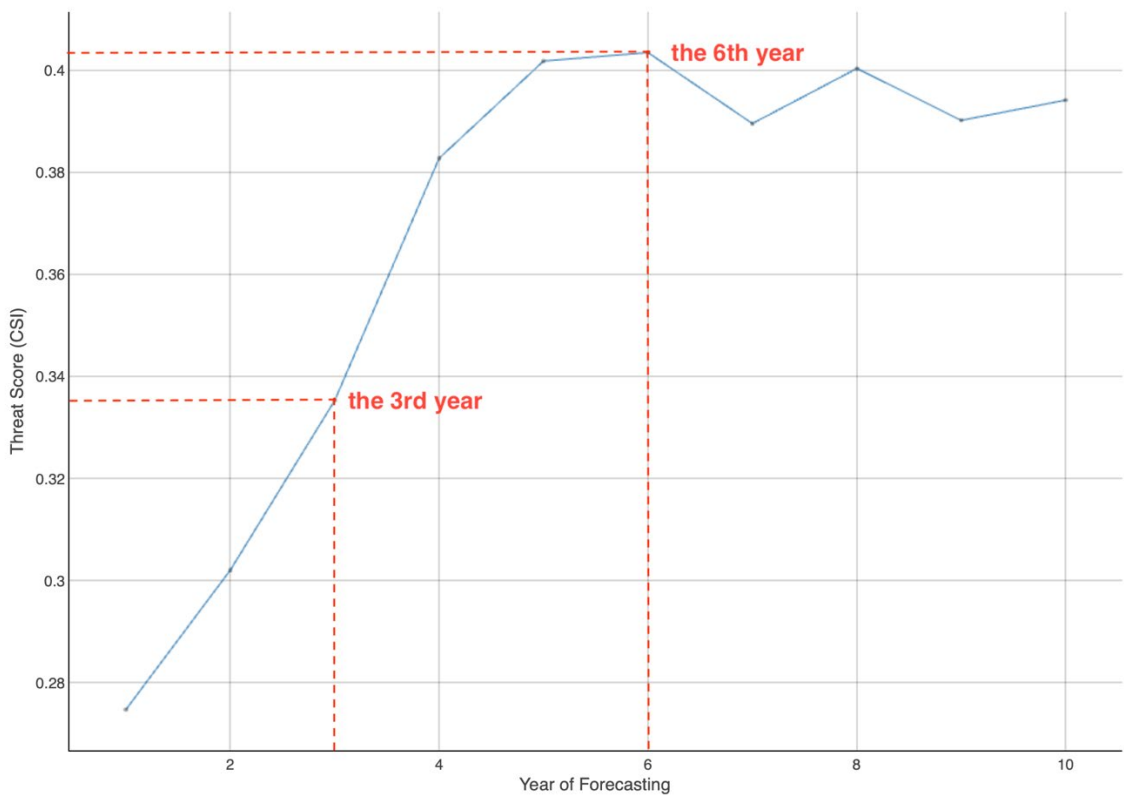

Figure 10: Threat score of prediction at different forecasting time points.

**Factors influencing future emergence**

To understand what factors influence the future emergence of a new MeSH concept, we ran logistic regression on the entire dataset. The regression coefficients are reported in Table 13.

Table 13: Logistic regression results on the entire dataset (*N*=1306)

| Variables | Coeff. | Std. Error | z value | Pr(>|z|) |
|---|---|---|---|---|
| **(Intercept)** | -0.295 | 0.191 | -1.550 | 0.121 |
| **NarrowerTerm** | -0.011 | 0.158 | -0.067 | 0.946 |
| **Category_Organisms (B)** | -0.582 | 0.223 | -2.617 | 0.009** |
| **Category_Chemicals & Drugs (D)** | -0.896 | 0.198 | -4.517 | 0.000*** |
| **Category_Diseases (C)** | -1.498 | 0.394 | -3.801 | 0.000*** |
| **ClinicalSignificance** | 2.215 | 0.137 | 16.202 | 0.000*** |
| **Signif. codes:  * p < 0.05. ** p <0.01. *** p<0.001** | | | | |

In terms of the effect of the three predictors, clinical significance has a significantly positive effect on the future emergence of a MeSH term (p<0.001) with a regression coefficient of 2.215. This means that the MeSH terms with clinical significance are more than nine times likely to emerge than those without ($e^{2.215} = 9.16$) holding other conditions constant. For topic categories, "Organisms," "Chemicals and Drugs," and "Diseases" have significantly negative effect on a topic's future emergence. The odds for topics in the category of "Organisms" to become emerging is only 0.55 times of that for other topics. Topics in the "Chemicals and Drugs" category have only 0.41 times the odds to emerge than that for other topics. Topics in "Diseases" have only 0.22 times the odds to emerge than that for other topics. The NarrowerTerm variable does not have a significant effect on topic emergence in the model.

## Discussion

### Summary of findings

Existing studies on emerging topic generally approach it from a publication perspective by retrieving a set of publications, and then using indicators from publications to identify or predict emerging topics. This study takes a topic perspective by tracing a set of newly added MeSH concepts in the biomedical domain and studying topic characteristics of the ones emerged versus those did not. In summary, we found topic characteristics do influence future popularity of a new MeSH:

- The broad topic category of a MeSH influences its future popularity. Organisms (B) category is more likely to be in the least popular quartile and less likely to be in the most popular quartile. Techniques and Equipment (E) category and Phenomena and Processes (G) category are more likely to be in the most popular quartile and less likely in the least popular quartile. Chemicals and Drugs (D) category is roughly evenly distributed among quartiles.

- MeSH concepts with clinical significance or usage are more popular than those without.
- MeSH concepts that have narrower concepts at the time of inclusion are less popular than those do not.

- Although many new concepts in Organisms (B) did not become popular, those pathogens for human attracted significantly more attention.

In general, we can summarize that the findings are consistent with the definition and mission of the biomedical field that more attention is paid to treating diseases and improving healthcare. New concepts that do not have immediate relevance with the mission may not attract much attention; therefore, they are not likely to become emerging topics.

In addition, we also identified four emerging trend patterns: "Emerged and Sustained", "Emerged not Sustained", "Emerged and Fluctuated", and "Not yet Emerged". We found among those emerged, a majority were able to sustain (about 65%). However, there are still 20.87% that not sustained, and 14.05% fluctuated. The trend patterns improve our understanding of the emergence process. Instead of thinking it as a binary (emerging vs. non-emerging), there are differences in emergence patterns. We also showed the emergence patterns depend on topic categories. D, E, and G categories are more likely to emerge and sustain, while B category is not likely to emerge or sustain. The interest on G category seems to be more stable than that on D and E categories.

The predictive models using topic characteristics show the features are able to predict emerging topics to some extent. The prediction capability depends on the forecasting time points. The best CSI score was observed at the sixth year since a new MeSH is added. It should be noted that although the same calculation was used for the threat score as that in studies from the publication perspective, such as Klavans, Boyack and Murdick (2020), the values are not comparable. Studies from the publication perspective retrieve a set of data in a certain period of time and work on all topics in the period, while the topic perspective only focuses on new topics. The predictors in this study are also complementary to other predictors discussed in literature.

**Time lag in knowing clinical significance**

Out of all topic characteristics we examined, the clinical significance, although still a feature of the topic, requires observation from publications with the "Clinical Trial" type. Because of this, it may take time before we know the clinical significance of a new concept. This may be attributed to our lack of domain knowledge to ascertain the clinical relevance of a MeSH concept, but can also be due to the development of domain knowledge. For example, the clinical significance of a concept may not be known until further discoveries.

For this reason, we examined how long it takes before we know a new concept has clinical significance/usage. We calculated the time interval between the year when a MeSH concept was added and the year when the first "Clinical Trial" article indexed by the new MeSH was published. We queried all 1279 MeSH in PubMed for the time difference. The overall time interval distribution is shown in Figure 11:

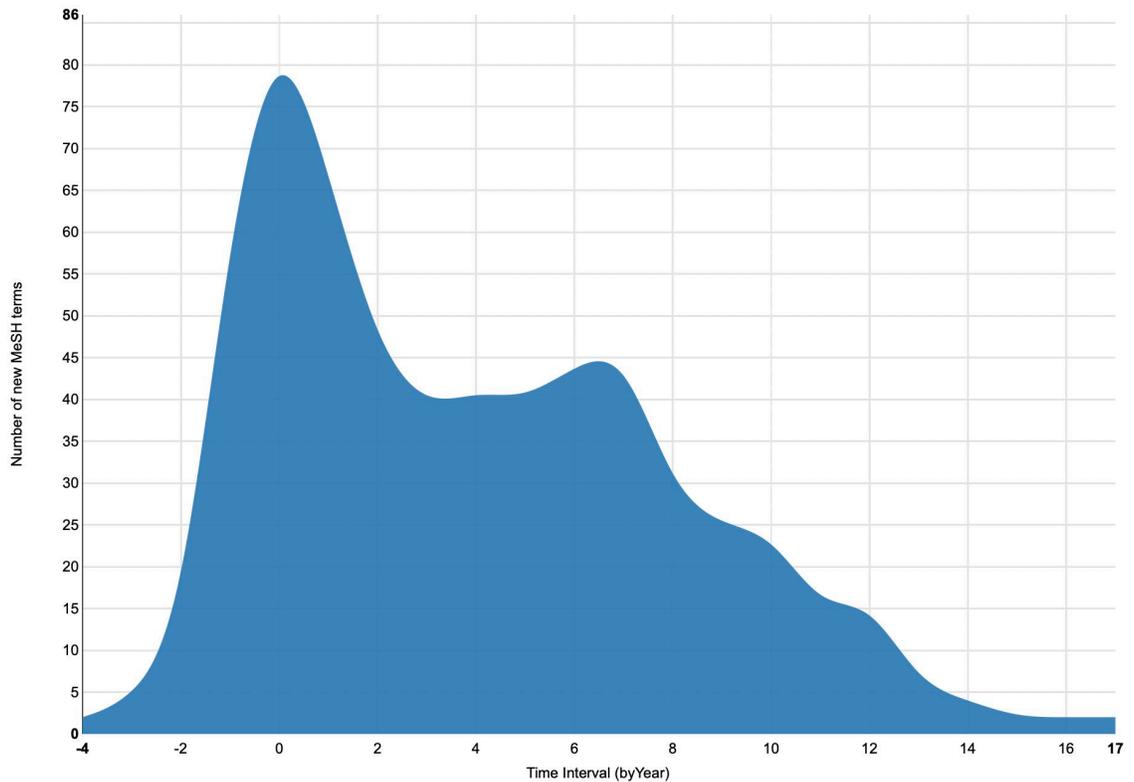

Figure 11: Time interval distribution between the years that a new MeSH was added and knowing its clinical significance ("Publication Year of the first clinical trial article indexed with the MeSH" - "Year Added of a MeSH").

Table 14 categorizes the interval into five stages. The first stage ranging from -4 to 0, meaning the clinical significance of a MeSH term would be known at the time of its inclusion or before its inclusion, because the first "Clinical Trial" article was published before the year that MeSH was added or in the same year. There are 27% of the MeSH with clinical significance belonging to this stage. The second stage indicates that we would know the clinical significance of a MeSH term within four years after its addition, accounting for 31.4% of all MeSH with clinical significance. Therefore, according to Table 14, within four years, we would know 58.5% of the MeSH terms with clinical significance. Within eight years, we would know 84.4% of them with clinical significance. Within twelve years, we would know the clinical significance of 97.4% of them.

Table 14: Time lag in knowing clinical significance (*N*=614).

| Stages | Time Lag (years) | Number of New MeSH Terms | Percentage | Cumulative Percentage |
|---|---|---|---|---|
| **The 1st Stage** | from -4 to 0 | 80 | 27% | 27% |
| **The 2nd Stage** | from 1 to 4 | 238 | 31.4% | 58.5% |
| **The 3rd Stage** | from 5 to 8 | 171 | 25.9% | 84.4% |
| **The 4th Stage** | from 9 to 12 | 93 | 13% | 97.4% |
| **The 5th Stage** | from 13 to 17 | 32 | 2.6% | 100% |

It is possible with the assistance of a domain expert, we may be able to know clinical significance of a MeSH even earlier, but the above table is based on indicators from publications.

**Contributions**

There are several contributions of this study:
- First of all, the study differs from the literature in studying emerging topics by taking a topic perspective. The characteristics of the topics are inherent features of topics and rooted in domain knowledge. One of the advantages of the topic perspective is we can not only study the emerged ones, but also the ones failed to emerge. This aspect is largely missing in existing research.
- The topic characteristics identified in this study explain why some topics emerged in the biomedical domain, while others did not. Factors are related to characteristics of the topics and are domain-specific. This study portraits a clearer picture of topics that emerged in the biomedical domain and their characteristics. By adopting a domain-specific approach, we are more able to reveal the specifics of information phenomena in a domain. This echoes the advocate for domain analytics (Hjørland, 2004). In fact, the topic characteristics that influence their emergence show close relationship with the mission of the biomedical domain. This points out that future studies on emerging topics should also consider the characteristics of the domain, which is generally neglected in existing studies.
- In addition, this study also identifies different trend patterns for emergence that describe what could happen after emergence.

**Limitations**

Several limitations of the study need to be acknowledged:

- First, although MeSH is a manually maintained controlled vocabulary for the biomedical field, there can be shift of meaning in a MeSH term over time (McCray & Lee, 2013). This may add complexity on whether the same MeSH term still refers to the same concept, and whether the scope of the concept changes over time.
- It should be pointed out that new concepts first appear in free text before they are added to controlled vocabularies. Delays in adoption are expected.
- To measure clinical significance of a topic, we relied on whether a MeSH term is assigned to at least one article with the type of "Clinical Trial". However, there are different phases of clinical trials, such as phase I, II, III, and IV. In addition, there are different types of clinical trials, such as adaptive clinical trial, controlled clinical trial, etc. This study did not differentiate the phases and types. It would be interesting to further explore the effect of those on topic emergence. However, there are also practical concerns. First, differentiating the phases may cause delay in measuring the variable. Finding the earliest clinical trial article shortens the time lag that has practical implications for predicting topic emergence. Second, fewer articles are indexed with those more specific publication types, which may cause a data sparsity issue.
- This study relied on PubMed for the scope of the biomedical field, which employs a Literature Selection Technical Review Committee (LSTRC) to review journal selection (https://www.nlm.nih.gov/lstrc/jsel.html). Despite this, it is difficult to clearly draw the boundary of a field, which may lead to debate on the scope.
- The study focuses on topic characteristics that complement other features discussed in literature. However, we did not carry out experiments in combining predictors for best prediction results because that is not the focus of the study.

**Conclusion**

This study traced a set of newly added MeSH from 2001 to 2010 and studied the topic characteristics that influence their emergence. We found topic characteristics, including topic categories, clinical significance and if a topic has narrower terms at the time of inclusion, influence its potential for emergence. In general, topics that are more relevant to the mission of the biomedical domain are more likely to become popular. The topic perspective does not assess what qualifies a topic for emerging (i.e. novelty, growth, impact, etc), but focuses on what topic characteristics drive a topic to emerge. This differs from the publication perspective that focuses on how a topic is discussed or cited in publications. We think the topic perspective is more fundamental because the characteristics of a topic may determine how much attention it will attract and how much impact it will generate. Different emergence trend patterns are also identified. Predictive analysis using topic characteristics for emerging topic prediction is carried out. This study advocates for more attention to the topic and domain characteristics when studying emerging topics, as well as considering economic, social, and environmental impact.

## Acknowledgements

This study extends a previous pilot work published in the proceedings of iConference 2020 (Lu, 2020) by increasing the sample size and adding more in-depth analysis and prediction. The authors appreciate the constructive comments from the anonymous reviewers that definitely help improve the manuscript.

## Reference

Asooja, K., Bordea, G., Vulcu, G., & Buitelaar, P. (2016). Forecasting Emerging Trends from Scientific Literature. In *Proceedings of Tenth International Conference on Language Resources and Evaluation (LREC'16)* (pp. 417–420). Retrieved from http://saffron.insight-centre.org/

Babko-Malaya, O., Seidel, A., Hunter, D., HandUber, J. C., Torrelli, M., & Barlos, F. (2015). Forecasting Technology Emergence from Metadata and Language of Scientific Publications and Patents. In *ISSI*.

Balogh SG, Zagyva D, Pollner P, Palla G (2019) Time evolution of the hierarchical networks between PubMed MeSH terms. *PLoS ONE 14*(8): e0220648. https://doi.org/10.1371/journal.pone.0220648

Carley, S. F., Newman, N. C., Porter, A. L., & Garner, J. G. (2018). An indicator of technical emergence. *Scientometrics, 115*(1), 35–49. https://doi.org/10.1007/s11192-018-2654-5

Chen, C. (2006). CiteSpace II: Detecting and Visualizing Emerging Trends and Transient Patterns in Scientific Literature. *Journal of the American Society for Information Science and Technology*, *57*(3), 359–377. https://doi.org/10.1002/asi

Cozzens, S., Gatchair, S., Kang, J., Kim, K. S., Lee, H. J., Ordóñez, G., & Porter, A. (2010). Emerging technologies: Quantitative identification and measurement. *Technology Analysis and Strategic Management*, *22*(3), 361–376. https://doi.org/10.1080/09537321003647396

Glänzel, W., & Thijs, B. (2012). Using 'core documents' for detecting and labelling new emerging topics. *Scientometrics, 91*(2), 399–416. https://doi.org/10.1007/s11192-011-0591-7

Guo, H., Weingart, S., & Börner, K. (2011). Mixed-indicators model for identifying emerging research areas. *Scientometrics*, *89*(1), 421–435. https://doi.org/10.1007/s11192-011-0433-7

Hjørland, B. (2004). Domain analysis: A socio-cognitive orientation for information science research. *Bulletin of the American Society for Information Science and Technology, 30*(3), 17-21.

Japkowicz, N. (2000). Learning from imbalanced data sets: a comparison of various strategies. In *AAAI workshop on learning from imbalanced data sets* (Vol. 68, pp. 10-15). AAAI Press Menlo Park, CA.

Kastrin, A., & Hristovski, D. (2019). Disentangling the evolution of MEDLINE bibliographic database: A complex network perspective. *Journal of Biomedical Informatics, 89(January)*, 101–113. https://doi.org/10.1016/j.jbi.2018.11.014

Klavans, R., Boyack, K. W., & Murdick, D. A. (2020). A novel approach to predicting exceptional growth in research. *Plos one*, *15*(9), e0239177.

Kyebambe, M. N., Cheng, G., Huang, Y., He, C., & Zhang, Z. (2017). Forecasting emerging technologies: A supervised learning approach through patent analysis. *Technological Forecasting and Social Change*, *125*(November 2016), 236–244. https://doi.org/10.1016/j.techfore.2017.08.002

Lee, C., Kwon, O., Kim, M., & Kwon, D. (2018). Early identification of emerging technologies: A machine learning approach using multiple patent indicators. *Technological Forecasting and Social Change*, *127*(October 2017), 291–303. https://doi.org/10.1016/j.techfore.2017.10.002

Lu, K. (2020). What kind of research topics emerged in the biomedical domain?: A perspective from newly added subject terms in a thesaurus. *iConference 2020 Proceedings*. Borås, Sweden.

McCray, A. T., & Lee, K. (2013). Taxonomic Change as a Reflection of Progress in a Scientific Discipline. In Evolution of Semantic Systems (pp. 189–208). Springer Berlin Heidelberg. https://doi.org/10.1007/978-3-642-34997-3_10

McKeown, K., Daume III, H., Chaturvedi, S., Paparrizos, J., Thadani, K., Barrio, P., ... & Teufel, S. (2016). Predicting the impact of scientific concepts using full-text features. *Journal of the Association for Information Science and Technology*, *67*(11), 2684-2696.

Moerchen, F., Fradkin, D., Dejori, M., & Wachmann, B. (2008). Emerging trend prediction in biomedical literature. *AMIA ... Annual Symposium Proceedings / AMIA Symposium. AMIA Symposium*, 485–489.

Ohniwa, R. L., Hibino, A., & Takeyasu, K. (2010). Trends in research foci in life science fields over the last 30 years monitored by emerging topics. *Scientometrics*, *85*(1), 111–127. https://doi.org/10.1007/s11192-010-0252-2

Porter, A. L., Garner, J., Carley, S. F., & Newman, N. C. (2019). Emergence scoring to identify frontier R&D topics and key players. *Technological Forecasting and Social Change, 146*, 628-643.

Price, D. J. D. S. (1965). Networks of scientific papers. *Science*, 510-515.

Upham, S. P., & Small, H. (2010). Emerging research fronts in science and technology: Patterns of new knowledge development. *Scientometrics*, *83*(1), 15–38. https://doi.org/10.1007/s11192-009-0051-9

Ranaei, S., & Suominen, A. (2017, July). Using machine learning approaches to identify emergence: Case of vehicle related patent data. In *2017 Portland International Conference on Management of Engineering and Technology (Picmet)* (pp. 1-8). IEEE.

Ranaei, S., Suominen, A., Porter, A., & Carley, S. (2020). Evaluating technological emergence using text analytics: two case technologies and three approaches. *Scientometrics, 122*(1), 215-247.

Rotolo, D., Hicks, D., & Martin, B. R. (2015). What is an emerging technology? *Research Policy*, *44*(10), 1827–1843. https://doi.org/10.1016/j.respol.2015.06.006

Small, H. (1973). Co-citation in the scientific literature: A new measure of the relationship between two documents. *Journal of the American Society for information Science*, *24*(4), 265-269.

Small, H., Boyack, K. W., & Klavans, R. (2014). Identifying emerging topics in science and technology. *Research Policy*, *43*(8), 1450–1467. https://doi.org/10.1016/j.respol.2014.02.005

Svenonius, E. (1989). Design of controlled vocabularies. *Encyclopedia of library and information science*, *45*(suppl 10), 82-109.

Tu, Y. N., & Seng, J. L. (2012). Indices of novelty for emerging topic detection. *Information Processing and Management*, *48*(2), 303–325. https://doi.org/10.1016/j.ipm.2011.07.006

Tsatsaronis, G., Varlamis, I., Kanhabua, N., & Nørvåg, K. (2013, March). Temporal classifiers for predicting the expansion of medical subject headings. In International Conference on Intelligent Text Processing and Computational Linguistics (pp. 98-113). Springer, Berlin, Heidelberg.

Wang, Q. (2018). A bibliometric model for identifying emerging research topics. *Journal of the Association for Information Science and Technology*, *69*(2), 290–304. https://doi.org/10.1002/asi.23930

Xu, S., Hao, L., An, X., Pang, H., & Li, T. (2020). Review on emerging research topics with key-route main path analysis. *Scientometrics*, *122*(1), 607–624. https://doi.org/10.1007/s11192-019-03288-5

Xu, S., Hao, L., Yang, G., Lu, K., & An, X. (2021). A Topic Models based Framework for Detecting and Forecasting Emerging Technologies. *Technological Forecasting and Social Change*, *162,* 120366.